\documentclass{llncs}
\usepackage[numbers,sectionbib]{natbib}
\AtBeginDocument{}

\usepackage{xspace}
\usepackage{listings}
\usepackage{lstautogobble}
\usepackage{booktabs}
\usepackage{balance}
\usepackage{url}

\usepackage{amsmath}
\usepackage{amsfonts}
\usepackage[numbers]{natbib}

\usepackage{algorithm}
\usepackage[noend]{algpseudocode}

\newcommand{\term}[1]{\emph{#1\/}}

\newcommand{\langH}{$\mathcal{H}$\xspace}

\newcommand*{\defeq}{\mathrel{\vcenter{\baselineskip0.5ex \lineskiplimit0pt
                     \hbox{\scriptsize.}\hbox{\scriptsize.}}}%
                     =}

\newcommand{\unify}{\approx}
\newcommand{\negf}{\sim}
\newcommand{\prog}{\mathsf{P}}

\newcommand{\vars}{\mathit{vars}}
\newcommand{\dom}{\mathit{dom}}

\newcommand{\gen}[1]{#1'}

\makeatletter
\newcommand{\mathleft}{\@fleqntrue\@mathmargin\parindent}
\newcommand{\mathcenter}{\@fleqnfalse}
\makeatother

\usepackage{ntheorem}

\begin{document}
\title
  %[A higher-order removal method]
  %{A higher-order removal method for definitional higher-order logic programs with fully-applied predicates}
   {Predicate Specialization for Definitional Higher-order Logic Programs}
\author{Antonis Troumpoukis \inst{1} \and Angelos Charalambidis\inst{2}}
\institute{%
  Department of Informatics and Telecommunications,
  \\University of Athens, Athens, Greece
  \\\email{antru@di.uoa.gr}
  \and
  Institute of Informatics and Telecommunications,
  \\ NCSR ``Demokritos'', Athens, Greece
  \\\email{acharal@iit.demokritos.gr}
}

\maketitle

\begin{abstract}
Higher-order logic programming is an interesting extension of traditional logic
programming that allows predicates to appear as arguments and variables to
be used where predicates typically occur. Higher-order characteristics are indeed desirable
but on the other hand they are also usually more expensive to support. In this paper we
propose a program specialization technique based on partial evaluation that can be
applied to a modest but useful class of higher-order logic programs and can transform
them into first-order programs without introducing additional data structures.
% The technique inspects the program for potential higher-order call bindings
% and creates appropriate specialized first-order clauses.
The resulting first-order programs can be executed
by conventional logic programming interpreters and benefit from other
optimizations that might be available. We provide an implementation and experimental
results that suggest the efficiency of the transformation.
\end{abstract}

%%%%%%%%%%%%%%%%%%%%%%%%%%%%%%%%%%%%%%%%%%%%%%%%%%%%%%%%%%%%%%%%%%%%%%%%%%%%%%%
%%%%%%%%%%%%%%%%%%%%%%%%%%%%%%%%%%%%%%%%%%%%%%%%%%%%%%%%%%%%%%%%%%%%%%%%%%%%%%%

%%%
\section{Introduction}
\label{sec:intro}
%%%

%Functional programming is traditionally correlated with the higher-order programmming
%paradigm and its success is often attributed on supporting higher-order
%constructs.

Higher-order logic programming has been long studied as an interesting extension
of traditional first-order logic programming and various approaches exist with
different features and semantics~\cite{hopes,hilog,lambdaprolog}.
Typically, higher-order logic programs are allowed to define predicates that accept
other predicates as arguments and variables can appear in places where predicate
constants typically occur. Higher-order logic programs enjoy similar merits as
their functional counterparts. The support of higher-order features however,
usually comes with a price, and the efficient implementation in either logic or
functional programming is a non-straightforward task.

The use of higher-order constructs is a standard feature in
every functional language in contrast to the logic programming languages.
As a result, there exists a plethora of optimizations that target specifically the
efficient implementation of such features.
A popular direction is to remove higher-order structures altogether by transforming
higher-order programs into equivalent first-order ones, with the hope that the
execution of the latter will be much more efficient. Reynolds, in his seminal
paper~\cite{Reynolds72}, proposed a defunctionalization algorithm that
is complete, i.e. it succeeds to remove all higher-order parameters from an arbitrary
functional program. There is however a tradeoff; his algorithm requires the introduction
of data structures in order to compensate for the inherent loss of expressivity~\cite{jones}.
Other approaches \cite{ChinD96,MitchellR09,Nelan92} have been proposed that do
not use data structures, but share the limitation that are not complete.

In the logic programming context there exist many transformation algorithms
with the purpose of creating more efficient programs.
Partial evaluation algorithms~\cite{Gallagher93,LloydS91,Leuschel98},
for example, can be used to obtain a more efficient program by iteratively
unfolding logic clauses. Most of the proposals, however, focus on first-order logic programs.
Proposals that can be applied to higher-order programs are limited. The prominent technique
that targets higher-order logic programs proposed in~\cite{warren1981higher, hilog}
and adopted from Hilog. It employs the Reynolds' defunctionalization adapted for
logic programs. As a consequence it naturally suffers from the same shortcomings
of the original technique: the resulting programs are not natural and the
conventional logic programming interpreters fail to identify potential
optimizations without specialized tuning~\cite{SagonasW95}.

In this paper, we propose a partial evaluation technique that can be applied
to higher-order logic programs. The technique propagates only higher-order arguments
and avoids to change the structure of the original program. Moreover, it differs
from Reynolds' style defunctionalization approaches as it does not rely on any type of
data structures. As a result, the technique will only guarantee to remove the
higher-order arguments in a well-defined subset of higher-order logic programs.
The main contributions of the present paper are the following:
\begin{enumerate}
\item We propose a technique based on the abstract framework of partial evaluation
      that targets higher-order arguments. We have identified a well-defined fragment
      of higher-order logic programming that the technique terminates and produces
      a logic program without higher-order arguments.
\item We provide an implementation of the proposed technique and we experimentally
      assess its performance. We also compare it with the Reynolds' defunctionalization
      implemented in Hilog. Moreover, we experiment with the ability of
      combining this technique with the well-known tabling optimization.
\end{enumerate}

The rest of the paper is organized as follows. In Section~\ref{sec:example}
we give an intuitive overview of our method using a simple example. In
Section~\ref{sec:lang} we formally define the fragment of the higher-order logic
programs we will use. Section~\ref{sec:partial} describes the abstract framework
of partial evaluation and Section~\ref{sec:method} introduces the details
of our method. Section~\ref{sec:implementation} discusses some implementation
issues and Section~\ref{sec:experiments} discusses the performance of our
transformation on various experiments. Lastly, we compare our method with
related approaches in Section~\ref{sec:related} and we conclude the paper with
possible future work.

%%%%%%%%%%%%%%%%%%%%%%%%%%%%%%%%%%%%%%%%%%%%%%%%%%%%%%%%%%%%%%%%%%%%%%%%%%%%%%%
%%%%%%%%%%%%%%%%%%%%%%%%%%%%%%%%%%%%%%%%%%%%%%%%%%%%%%%%%%%%%%%%%%%%%%%%%%%%%%%

%%
\section{A Simple Example}
\label{sec:example}

We start with an introductory example so as to give an informal description of our technique.
We borrow an example from the area of knowledge representation which deals with the
expression of user preferences~\cite{holppref}.

The following program selects the most preferred tuples {\tt T} out of a given
unary relation {\tt R}, based on a binary preference predicate {\tt P}. The preference
predicate given two tuples it succeeds if the first tuple is more preferred than the second.
\begin{lstlisting}
  winnow(P,R,T) :- R(T), not bypassed(P,R,T).
  bypassed(P,R,T) :- R(Z), P(Z,T).
\end{lstlisting}
The program contains \term{predicate variables} (for example, {\tt P} and {\tt R}),
that is variables that can occur in places where predicates typically occur.

Assume that we have a unary predicate \lstinline{movie} which corresponds to a
relation of movies and a binary predicate \lstinline{pref} which given two movies
succeeds if the first argument has a higher ranking than the second one.
Now, suppose that we issue the query:
\begin{lstlisting}
  ?- winnow(pref,movie,T).
\end{lstlisting}
We expect as answers the most ``preferred'' movies, that is all movies with the
highest ranking.

In the following, we will show how we can create a first-order version of the original
program specialized for this specific query.
Notice that the atom \lstinline{winnow(pref,movie,T)}, that makes up our
given query, does not contain any free predicate variables, but on the contrary,
all of its  predicate variables are substituted with predicate names. Therefore,
we can specialize every program clause that defines \lstinline{winnow} by
substituting its predicate variables with the corresponding predicate names.
By doing so, we get a program where our query yields the same results as to
those in the original program:
\begin{lstlisting}
  winnow(pref,movie,T) :- movie(T), not bypassed(pref,movie,T).
  bypassed(P,R,T) :- R(Z), P(Z,T).
\end{lstlisting}
We can continue this specialization process by observing that in the body of
this newly constructed clause there exists the atom
\lstinline{bypassed(pref,movie,T)}, in which all predicate variables
are again substituted with predicate names. Therefore, we can specialize
the second clause of the program accordingly:
\begin{lstlisting}
  winnow(pref,movie,T) :- movie(T), not bypassed(pref,movie,T).
  bypassed(pref,movie,T) :- movie(Z), pref(Z,T).
\end{lstlisting}
There are no more predicate specializations to be performed and the transformation stops.
Notice that the resulting program does not contain any predicate variables, but
it is not a valid first-order one. Therefore, we have to perform a simple rewriting
in order to remove all unnecessary predicate names that appear as arguments.
\begin{lstlisting}
  winnow1(T) :- movie(T), not bypassed2(T).
  bypassed2(T) :- movie(Z), pref(Z,T).
\end{lstlisting}
Due to this renaming process, instead of the initial query,
the user now has to issue the query \lstinline{?- winnow1(T).}
Comparing the final first-order program with the original one it is easy to observe
that no additional data structures were introduced during the first-order
transformation, a characteristic that leads to performance improvement
(ref. Section~\ref{sec:experiments}).

This technique, however, cannot be applied in every higher-order logic program.
Notice that the resulting program of the previous example does not contain any predicate
variables. This holds due to the fact that in the original program, every
predicate variable that appears in the body of a clause it also appears in the
head of this clause. By restricting ourselves to programs that have this property
we ensure that the transformation outputs a first-order program.
Moreover, the transformation in this example terminates because the set of the
\term{specialization atoms} (ie. \lstinline{winnow(pref,movie,T)} and
\lstinline{bypassed(pref,movie,T)}) is finite, which is not the
case in every higher-order logic program. To solve this, we need to keep the set
specialization atoms finite. This is achieved in two ways.
Firstly, we ignore all first-order arguments in every specialization atom,
meaning that in a query of the form \lstinline{?- winnow(pref,movie,m_001)}, we
will specialize the program with respect to the atom \lstinline{winnow(pref,movie,T)}.
Secondly, we impose one more program restriction;
we focus in programs where the higher-order arguments are either variables or
predicate names. Since the set of all predicate names is finite and since we
ignore all first-order values, the set of specialization atoms is also finite and
as a result the algorithm is ensured to terminate.

\section{Higher-order Logic Programs}
\label{sec:lang}

In this section we define the higher-order language of our interest.
We begin with the syntax of the language \langH we use throughout the paper.
\langH is based on a simple type system with two base types:
$o$, the boolean domain, and $\iota$, the domain of data objects.
The composite types are partitioned into three classes:
\term{functional} (assigned to function symbols),
\term{predicate} (assigned to predicate symbols) and
\term{argument} (assigned to parameters of predicates).

\begin{definition}
  A type can either be functional, argument, or predicate, denoted by
    $\sigma, \rho$ and $\pi$ respectively and defined as:
  \[
  \begin{array}{lclcl}
    \sigma & \defeq & \iota & | & ( \iota \to \sigma ) \\
    \pi    & \defeq & o     & | & ( \rho \to \pi ) \\
    \rho   & \defeq & \iota & | & \pi \\
  \end{array}
  \]
\end{definition}
\begin{definition}
%The {\em alphabet} of the higher-order language \langH consists of the following:
The \term{alphabet} of the language \langH consists of the following:
\begin{enumerate}
  \item Predicate variables of every predicate type $\pi$
        (denoted by capital letters such as $\mathsf{P,Q,R,\ldots}$).
  \item Individual variables of type $\iota$
        (denoted by capital letters such as $\mathsf{X,Y,Z,\ldots}$).
  \item Predicate constants of every predicate type $\pi$
        (denoted by lowercase letters such as $\mathsf{p,q,r,\ldots}$).
  \item Individual constants of type $\iota$
        (denoted by lowercase letters such as $\mathsf{a,b,c,\ldots}$).
  \item Function symbols of every functional type $\sigma \neq \iota$
        (denoted by lowercase letters such as $\mathsf{f,g,h,\ldots}$).
  \item The inverse implication constant $\gets$, the negation constant $\negf$,
        the comma, the left and right parentheses, and the equality constant
        $\unify$ for comparing terms of type $\iota$.
\end{enumerate}
\end{definition}

The set consisting of the predicate variables and the individual variables of
\langH will be called the set of \term{argument variables} of \langH. Argument
variables will be usually denoted by $\mathsf{V}$ and its subscripted versions.

\begin{definition}
%The set of \term{expressions} of the higher-order language \langH is defined as follows:
The set of \term{expressions} of \langH is defined as follows:
\begin{itemize}
  \item Every predicate variable (resp. predicate constant) of type $\pi$ is an
        expression of type $\pi$; every individual variable (resp. individual constant)
        of type $\iota$ is an expression of type $\iota$;
  \item if $\mathsf{f}$ is an $n$-ary function symbol and $\mathsf{E}_1, \ldots, \mathsf{E}_n$
        are expressions of type $\iota$ then $(\mathsf{f}\ \mathsf{E}_1\cdots\mathsf{E}_n)$ is
        an expression of type $\iota$;
  \item if $\mathsf{E}$ is an expression of type $\rho_1 \rightarrow \cdots \rho_n \rightarrow o$ and
        $\mathsf{E}_i$ an expression of type $\rho_i$ for $i \in \{ 1, \ldots, n\}$ then
        $(\mathsf{E}\ \mathsf{E}_1\ \cdots\ \mathsf{E}_n)$ is an expression of type $o$.
  \item if $\mathsf{E}_1, \mathsf{E}_2$ are expressions of type $\iota$,
        then $(\mathsf{E}_1 \unify \mathsf{E}_2)$ is an expression of type $o$.
\end{itemize}
\end{definition}

We will omit parentheses when no confusion arises.
% To denote that an expression $\mathsf{E}$ has type $\rho$ we will often write $\mathsf{E}:\rho$.
Expressions of type $o$ will often be referred to as \term{atoms}.
We write $\vars(\mathsf{E})$ to denote the set of all variables in $\mathsf{E}$.
We say that $\mathsf{E}_i$ is the $i$-th argument of an atom $\mathsf{E}\ \mathsf{E}_1\ \cdots\ \mathsf{E}_n$.
A \emph{ground expression} $\mathsf{E}$ is an expression where $\vars(\mathsf{E})$
is the empty set.

\begin{definition}
\label{def:langH}
A \term{clause} is a formula
\[ \mathsf{p}\ \mathsf{V}_1 \cdots \mathsf{V}_n \gets \mathsf{L}_1, \ldots, \mathsf{L}_m, \negf\mathsf{L}_{m+1},\ldots,\negf\mathsf{L}_{m+k}\]
where $\mathsf{p}$ is a predicate constant of type $\rho_1 \to \cdots \to \rho_n \to o$,
$\mathsf{V}_1,\ldots,\mathsf{V}_n$ are distinct variables of types $\rho_1,\ldots,\rho_n$
respectively, and $\mathsf{L}_1,\ldots,\mathsf{L}_{m+k}$ are expressions of type $o$,
such that every predicate argument of $\mathsf{L}_i$ is either variable or ground.
%The term $\mathsf{p}\ \mathsf{V}_1 \cdots \mathsf{V}_n$ is called the \term{head}
%of the clause, the variables $\mathsf{V}_1, \ldots, \mathsf{V}_n$ are the
%\term{formal parameters} of the clause and the conjunction
%$\mathsf{L}_1, \ldots, \mathsf{L}_m, \negf\mathsf{L}_{m+1},\ldots,\negf\mathsf{L}_{m+k}$
%is its \term{body}.
A \term{program} $\prog$ of the higher-order language \langH is a finite set of program clauses.
\end{definition}

The syntax of programs given in Definition~\ref{def:langH} differs slightly from
the usual Prolog-like syntax that we have used in Section~\ref{sec:example}. However,
one can easily verify that we can rewrite every program from the former syntax
to the latter. For instance, we could use the constant $\unify$ in order to
eliminate individual constants that appear in the head of a clause that uses
the Prolog-like syntax.

\begin{example}
\label{ex:syntax}
Consider the following program in Prolog-like syntax, in which we have three
predicate definitions, namely
$\mathtt{p : \iota \to o}$,
$\mathtt{q : \iota \to \iota \to o}$, and
$\mathtt{r} : (\iota \to o) \to (\iota \to o) \to (\iota \to \iota) \to o$.
\begin{lstlisting}
  p(a).
  q(X,X).
  r(P,Q,f(X)) :- P(X),Q(Y).
\end{lstlisting}
In our more formal notation, these clauses can be rewritten as:
\mathleft
\[
\begin{array}{@{}l}
  \mbox{\tt p X     $\gets$ (X $\unify$ a).}\\
  \mbox{\tt q X Y   $\gets$ (X $\unify$ Y).}\\
  \mbox{\tt r P Q Z $\gets$ (Z $\unify$ f(X)), (P X), (Q Y).}
\end{array}
\]
\mathcenter
Notice that all clauses are now valid \langH clauses.
\end{example}

Notice that in a \langH program, all arguments of predicate type
are either variables or predicate names, which as dicussed in Section~\ref{sec:example}
leads to the termination of our technique.
However, in a \langH program all head predicate variables to be distinct.
That implies that checking for equality between predicates (higher-order unification) is forbidden.
In other words, the higher-order parameters can be used in ways similar to functional programming, namely
either invoked or passed as arguments. We decided to impose this restriction
because equality between predicates is treated differently in various higher-order
languages~\cite{hopes,hilog,lambdaprolog}.
Moreover, in Section~\ref{sec:example},
we briefly discussed that the reason why our technique can produce a first-order
program is due to the following property:

\begin{definition} \label{def:langDef}
A clause will be called \term{definitional} iff every predicate variable that
appears in the body appears also as a formal parameter of the clause.
A \term{definitional program} is a finite set of definitional clauses.
\end{definition}

\begin{example}
Consider the following program in Prolog-like syntax:
\begin{lstlisting}
  p(Q,Q) :- Q(a).
  q(X) :- R(a,X).
\end{lstlisting}
This program does not belong to our fragment, because the first clause is a non-\langH
clause and the second clause is a non-definitional clause.
Regarding the first clause, the predicate variable {\tt Q} appears twice in
the head, therefore the formal parameters are not distinct.
Regarding the second clause, the predicate variable {\tt R} that appears in
the body, does not appear in the head of the clause.
\end{example}

We extend the well-known notion of substitution to apply to \langH programs.

\begin{definition}
A \term{substitution} $\theta$ is a finite set $\{ \mathsf{V}_1/\mathsf{E}_1, \ldots,
\mathsf{V}_n/\mathsf{E}_n \}$ where the $\mathsf{V}_i$'s are different argument
variables and each $\mathsf{E}_i$ is a term having the same type as $\mathsf{V}_i$.
%Each element $\mathsf{V}_i/\mathsf{E}_i$ is called a \term{binding} for
%$\mathsf{V}_i$.
We write $\dom(\theta) = \{ \mathsf{V}_1, \ldots, \mathsf{V}_n\}$ to denote the domain of $\theta$.
%and $\codom(\theta) = \{ \mathsf{E}_1, \ldots,
%\mathsf{E}_n\}$ to denote the co-domain of $\theta$. %A substitution is called
%\term{ground} if every expression $\mathsf{E}_i$ is ground. The substitution
%corresponding to the empty set will be called the identity substitution denoted
%as $\epsilon$.
\end{definition}

\begin{definition}
Let $\theta$ be a substitution and $\mathsf{E}$ be an expression.
Then, $\mathsf{E}\theta$ is an expression obtained from $\mathsf{E}$ as follows:
\begin{itemize}
  \item $\mathsf{E}\theta = \mathsf{E}$ if $\mathsf{E}$ is a predicate constant or individual constant;
  \item $\mathsf{V}\theta = \theta(\mathsf{V})$ if $\mathsf{V} \in \dom(\theta)$;
        otherwise, $\mathsf{V}\theta = \mathsf{V}$;
  \item $(\mathsf{f}\ \mathsf{E}_1\cdots\mathsf{E}_n)\theta = (\mathsf{f}\ \mathsf{E}_1\theta\cdots\mathsf{E}_n\theta)$;
  \item $(\mathsf{E}\ \mathsf{E}_1\ \cdots\ \mathsf{E}_n)\theta = (\mathsf{E}\theta\ \mathsf{E}_1\theta\ \cdots\ \mathsf{E}_n\theta)$;
  \item $(\mathsf{E}_1\approx \mathsf{E}_2)\theta = (\mathsf{E}_1\theta\approx \mathsf{E}_2\theta)$;
  %\item $(\negf \mathsf{E})\theta = (\negf \mathsf{E}\theta)$;
  \item $(\mathsf{L}_1, \ldots, \mathsf{L}_m, \negf\mathsf{L}_{m+1},\ldots,\negf\mathsf{L}_n)\theta =
         (\mathsf{L}_1\theta, \ldots, \mathsf{L}_m\theta, \negf(\mathsf{L}_{m+1}\theta),\ldots,\negf(\mathsf{L}_n\theta))$.
\end{itemize}
\end{definition}
Let $\theta$ be a substitution and $\mathsf{E}$ an expression. Then, $\mathsf{E}\theta$
is called an \emph{instance} of $\mathsf{E}$.

\section{Partial Evaluation of Logic Programs}
\label{sec:partial}

\term{Partial evaluation}~\cite{jonesetal93} is a program optimization that
specializes a given program according to a specific set of input data, such that
the new program is more efficient than the original and both programs behave in
the same way according to the given data.
In the context of logic programming~\cite{Gallagher93,LloydS91,Leuschel98},
a partial evaluation algorithm takes a program $\mathsf{P}$ and a goal $\mathsf{G}$
and produces a new program $\mathsf{P}'$ such that
$\mathsf{P} \cup \{\mathsf{G}\}$ and $\mathsf{P}' \cup \{\mathsf{G}\}$ are
semantically equivalent.
In Figure~\ref{alg:pareval} we illustrate a basic scheme that aims to describe
every partial evaluation algorithm in logic programming, which is based in
similar ones in the literature~\cite{Gallagher93,Leuschel98}.
Notice that this general algorithm depends on two operations, namely
\textsc{Unfold} and \textsc{Abstract}, which can be implemented differently
in several partial evaluation systems.

\begin{figure}[t]
  \begin{algorithmic}[1]
    \State \textbf{Input:} a program $\mathsf{P}$ and a goal $\mathsf{G}$
    \State \textbf{Output:} a specialized program $\mathsf{P}'$
    \State $\mathsf{S} := \{ \mathsf{A} : \mathsf{A}
                              \mbox{ is an atom of } \mathsf{G} \}$
    \Repeat
      \State $\mathsf{S}' := \mathsf{S}$
      \State $\mathsf{P}' := $ \Call{Unfold}{$\mathsf{P},\mathsf{S}$}
      \State $\mathsf{S} := \mathsf{S} \cup
             \{ \mathsf{A} :\mathsf{A} \mbox{ is an atom that appears in a body
             of a clause in } \mathsf{P}' \}$
      \State $\mathsf{S} := $ \Call{Abstract}{$\mathsf{S}$}
    \Until {$\mathsf{S}' = \mathsf{S}$} (modulo variable renaming)
    \State \textbf{return} $\mathsf{P}'$
    \end{algorithmic}
  \caption{Basic Algorithm for Partial Evaluation.}
  \label{alg:pareval}
\end{figure}

Firstly, the algorithm uses an \term{unfolding} rule~\cite{Shepherdson92}
in order to construct a finite and possibly incomplete proof tree for
every atom in the set $\mathsf{S}$ and then creates a program $\prog'$ such that
every clause of it is constructed from all root-to-leaf derivations of these
proof trees. This part of the process is referred as the \term{local control}
of partial evaluation.
There are many possible unfolding rules, some of which being more useful for a
particular application than others. Examples include determinate,
leftmost non-determinate, loop-preventing or depth-bound unfolding
strategies~\cite{Gallagher93,Leuschel98}.
In some cases though, taking a simple approach which performs no unfolding
at all, or in other words by using \term{one-step unfolding strategy}, may
result in useful program optimizations.
In such a case, \textsc{Unfold} exports a program that is constructed by
finding the clauses that unify with each atom in $\mathsf{S}$ and then by
specializing these clauses accordingly, using simple variable substitutions.

Secondly, the algorithm uses an \textsc{Abstract} operation, which calculates a
finite \term{abstraction} of the set $\mathsf{S}$.
We say that $\mathsf{S'}$ is an \term{abstraction} of $\mathsf{S}$
if every atom of $\mathsf{S}$ is an instance of some atom in $\mathsf{S'}$,
and there does not exist two atoms in $\mathsf{S'}$ that have a common instance
in $\mathsf{S'}$. This operation is used to keep the size of the set of atoms
$\mathsf{S}$ finite, which will ensure the termination of the algorithm.
This part of the process is referred as the \term{global control} of partial
evaluation. Examples of abstraction operators include the use of a most specific
generalizer and a finite bound in the size of $\mathsf{S}$~\cite{Leuschel98},
or by exploiting a distinction between static and dynamic arguments
for every atom in $\mathsf{S}$~\cite{LeuschelVidal14}.

A partial evaluation algorithm should ensure termination in both levels of control.
Firstly, we have the \term{local termination problem},
which is the problem of the non-termination of the unfolding rule,
and the \term{global termination problem} which is the problem of the
non-termination of the iteration process (ie. the repeat loop in the algorithm).
As we stated earlier, the global termination problem is solved by keeping
the set $\mathsf{S}$ finite through a finite abstraction operation.
Regarding the local termination problem, one possible solution is ensuring that
all constructed proof trees are finite. The one-step unfolding rule is by
definition a strategy that can ensure local termination.

%%%%%%%%%%%%%%%%%%%%%%%%%%%%%%%%%%%%%%%%%%%%%%%%%%%%%%%%%%%%%%%%%%%%%%%%%%%%

% In order to discuss the basic points of partial evaluation in logic programming,
% we assume basic familiarity with SLDNF resolution of classical logic programming.
% In addition, we will need the following definition~\cite{Gallagher93,LloydS91}:
%
% \begin{definition}
%   Let $\mathsf{P}$ be a logic program and $\mathsf{S}$ be a set of atoms.
%   We say that $\mathsf{P}$ is \term{$\mathsf{S}$-closed} if every atom that
%   appears in $\mathsf{P}$ is an instance of some atom in $\mathsf{S}$.
%   Moreover, we say that $\mathsf{S}$ is \term{independent} if no two atoms
%   in $\mathsf{S}$ have a common instance in $\mathsf{S}$.
%   Finally, we say that $\mathcal{A'}$ is an \term{abstraction} of $\mathsf{S}$
%   if $\mathcal{A'}$ is independent and every atom of $\mathsf{S}$ is an instance
%   of some atom in $\mathcal{A'}$.
% \end{definition}

%%%%%%%%%%%%%%%%%%%%%%%%%%%%%%%%%%%%%%%%%%%%%%%%%%%%%%%%%%%%%%%%%%%%%%%%%%%%%%%
%%%%%%%%%%%%%%%%%%%%%%%%%%%%%%%%%%%%%%%%%%%%%%%%%%%%%%%%%%%%%%%%%%%%%%%%%%%%%%%

%%
\section{Predicate Specialization}
% \section{Predicate Specialization of Higher-Order Logic Programs}
\label{sec:method}
%%

% In this section we present the details of our method.

% \begin{definition}
% \label{def:skeleton}
% Let $\mathsf{A} = \mathsf{p}\ \mathsf{E}_1 \cdots \mathsf{E}_n$ be an atom.
% Then, the \term{skeleton} of $\mathsf{A}$, denoted as $\skel(\mathsf{A})$,
% is an expression of the form $\mathsf{p}\ \mathsf{E'}_1 \cdots \mathsf{E'}_n$
% s.t. for all $1 \le i \le n$:
% \begin{enumerate}
%   \item if $\mathsf{E}_i$ is of predicate type, then $\mathsf{E'}_i = \mathsf{E}_i$;
%   \item otherwise, $\mathsf{E'}_i = \mathsf{V}_i$, where $\mathsf{V}_i$ is a fresh variable
%     of the same type as of $\mathsf{E}_i$.
% \end{enumerate}
% \end{definition}

In the following, we define our technique using the standard framework of
partial evaluation (ref. Section~\ref{sec:partial}), by specifying its local
and global control strategies (namely \textsc{Unfold} and \textsc{Abstract}
operations). In particular, we will use a one-step unfolding
rule and an abstraction operation which generalizes all individual
(ie. non-predicate) arguments from all atoms of the partial evaluation.

\begin{definition}
\label{def:unfold}
Let $\prog$ be a program and $\mathsf{S}$ be a set of atoms.
Then,
\[
  \textsc{Unfold}(\prog,\mathsf{S}) =
  \left\{
    \mathsf{p}\ \mathsf{E}_1 \cdots \mathsf{E}_n \gets \mathsf{B}\theta :
    \begin{array}{l}
      (\mathsf{p}\ \mathsf{E}_1 \cdots \mathsf{E}_n)  \in \mathsf{S}, \\
      (\mathsf{p}\ \mathsf{V}_1 \cdots \mathsf{V}_n \gets \mathsf{B} ) \in \prog, \\
      \theta = \{\mathsf{V}_1/\mathsf{E}_1, \dots, \mathsf{V}_n/\mathsf{E}_n \}
    \end{array}
  \right\}
\]
\end{definition}
\begin{definition}
\label{def:abstract}
Let $\mathsf{S}$ be a set of atoms.
Then,
\[
  \textsc{Abstract}(\mathsf{S}) =
  \left\{
    \mathsf{p}\ \gen{\mathsf{E}_1} \cdots \gen{\mathsf{E}_n} :
    \begin{array}{l}
      (\mathsf{p}\ \mathsf{E}_1 \cdots \mathsf{E}_n)  \in \mathsf{S}
    \end{array}
  \right\}
\]
where $\gen{\mathsf{E}_i}=\mathsf{E}_i$ if $\mathsf{E}_i$ is of predicate type,
otherwise $\gen{\mathsf{E}_i} = \mathsf{V}_i$, where $\mathsf{V}_i$ is a fresh variable
of the same type as of $\mathsf{E}_i$.
\end{definition}

In the following, we will show some properties of our transformation.
Firstly we will need the following lemma:

\begin{lemma}
  Let $\prog$ be a program, $\mathsf{S}$ be a (possibly infinite) set of atoms. Then:
  \begin{enumerate}
    \item If $\mathsf{S}$ is finite, then $\textsc{Unfold}(\prog,\mathsf{S})$ is finite.
    \item $\textsc{Abstract}(\mathsf{S})$ is a finite abstraction of $\mathsf{S}$.
    \item If every element of $\mathsf{S}$ does not contain any free predicate variables,
          then every atom of $\textsc{Unfold}(\prog,\mathsf{S})$ does not contain
          any free predicate variables.
  \end{enumerate}
  \begin{proof}
    \begin{enumerate}
      \item Obvious from the construction of $\textsc{Unfold}(\prog,\mathsf{S})$.
      \item Every predicate argument of every atom that appears in $\prog$ is
            either a variable or a predicate name, therefore $\textsc{Abstract}(\mathsf{S})$
            is finite.
      \item Suppose that $\textsc{Unfold}(\prog,\mathsf{S})$ contains an atom
            $\mathsf{A}$ that contains a free predicate variable $\mathsf{V}$.
            If $\mathsf{A}$ appears in the head of a clause, then from the construction
            of $\textsc{Unfold}(\prog,\mathsf{S})$, $\mathsf{S}$ must contain $\mathsf{A}$.
            If $\mathsf{A}$ appears in the body of a clause, then since $\prog$
            is definitional, $\mathsf{V}$ also appears in the head of this clause.
            In any case, $\mathsf{S}$ must contain an atom that contains
            the free predicate variable $\mathsf{V}$.
    \end{enumerate}
  \end{proof}
\end{lemma}
The first part of the lemma ensures
local termination and the second part of the lemma ensures global termination.
The third part identifies that the transformation to first-order succeeds, provided
that the program belongs to our fragment and the initial goal does not contain
free higher-order variables. In the following corollaries, by $\Phi$
we denote the algorithm of Figure~\ref{alg:pareval} combined with the operations
in Definitions~\ref{def:unfold}~and~\ref{def:abstract} .

\begin{corollary}
  Let $\prog$ be a \langH program and $\mathsf{G}$ an goal.
  Then, the computation of $\Phi(\prog,\mathsf{G})$ terminates in a finite number
  of steps.
\end{corollary}

\begin{corollary}
  Let $\prog$ be a definitional program and $\mathsf{G}$ an goal
  that does not contain any free predicate variables.
  Then, the output of $\Phi(\prog,\mathsf{G})$ does not contain any free
  predicate variables.
\end{corollary}

The result of $\Phi$ is neither a valid \langH program since it contains predicate names as arguments in the heads,
nor a valid first-order program since some symbols appear both as arguments and
as predicate symbols. Therefore, we must apply a simple
\term{renaming}~\cite[Section~3]{Gallagher93} in order to construct a valid
first-order output.
In our case, at the end of the partial evaluation algorithm, every atom
$\mathsf{p}\ \mathsf{E}_1 \cdots \mathsf{E}_n$ of $\mathsf{S}$
is renamed into $\mathsf{p}'\ \mathsf{V}_1 \cdots \mathsf{V}_m$,
where $\mathsf{p}'$ is a fresh predicate symbol and
$\{\mathsf{V}_1 \cdots \mathsf{V}_m\} =
 \vars(\mathsf{p}\ \mathsf{E}_1 \cdots \mathsf{E}_n)$.
Moreover, all instances of every atom of $\mathsf{S}$ in the resulting program
are renamed accordingly.

%%%%%%%%%%%%%%%%%%%%%%%%%%%%%%%%%%%%%%%%%%%%%%%%%%%%%%%%%%%%%%%%%%%%%%%%%%%%%%%
%%%%%%%%%%%%%%%%%%%%%%%%%%%%%%%%%%%%%%%%%%%%%%%%%%%%%%%%%%%%%%%%%%%%%%%%%%%%%%%

%%
\section{Implementation}
\label{sec:implementation}

We have developed a prototype implementation\footnote{The implementation of the
transformation is open source and can be accessed at \url{http://bitbucket.org/antru/firstify}.}
of our predicate specialization technique.
Instead of developing a tailor-made higher-order language only for the purpose of
demonstrating the benefits of the transformation, we build upon an
existing higher-order logic programming language. The source programs in
have to be written in the higher-order language
Hilog~\cite{hilog}, a mature and well-known language with a stable implementation
within the XSB system~\cite{xsb}.

A feature that we need and is not supported in Hilog though,
is the use of types. Our algorithm needs types not only for deciding whether the
input program belongs to our fragment, but also for the abstraction operation
in Definition~\ref{def:abstract}. Since the process of extending Hilog with types
is outside of the scope of this paper, we assume that the input programs are
well-typed and accompanied with type annotations for all predicates that contain
predicate arguments.

The fragment that we discussed in Section~\ref{sec:lang} consists of programs
that the only elements that can appear as predicate arguments are variables and
predicate constants. However, most higher-order languages (and Hilog among them)
allow more complex expressions to appear as predicate arguments. One such
example is the use of \term{partial applications}, ie., the ability to
apply a predicate to only some of its arguments. Consider the following simple
program.
\begin{lstlisting}
  conj2(P,Q,X) :- P(X),Q(X).
  conj3(P,Q,R,X) :- conj2(P,conj2(Q,R),X).
\end{lstlisting}
In the second clause the expression {\tt conj2(Q,R)} is a partial application
where only the first two arguments are defined. A partial application effectively
produces a new relation and therefore typically occur in higher-order arguments.

In the implementation we are able to handle programs that make a limited use of
complex predicate expressions, as a syntactic sugar for our initial fragment.
In particular, we allow non-variable and non-constant predicate arguments in
an expression of the form
$\mathsf{p}\ \mathsf{E}_1 \cdots \mathsf{E}_n$
that appears in the body of a clause
$\mathsf{q}\ \mathsf{V}_1 \cdots \mathsf{V}_m \gets \mathsf{B}$
only if $\mathsf{p}$ and $\mathsf{q}$ do not belong in the same cycle in the
\term{predicate dependency graph}.~\footnote{An edge from the predicate
$\mathsf{p}$ to predicate $\mathsf{q}$ in the predicate dependency graph means
that there exists a clause that $\mathsf{p}$ appears in the head and $\mathsf{p}$
appears in the body of the same clause.}
The transformation in this case is also ensured to terminate (because due to the
form of the program all predicate variables of a predicate that depends on
itself have to be specialized only with predicate names and therefore the set of
all possible specialization atoms will remain finite). As we mentioned earlier,
this class of programs has the same expressive power as our initial fragment.
For example, the aforementioned logic program is equivalent to the following
program that does not use any partial applications.
\begin{lstlisting}
  conj2(P,Q,X) :- P(X),Q(X).
  conj31(P,Q,R,X) :- conj22(P,Q,R,X).
  conj22(P,Q,R,X) :- P(X),conj2(Q,R,X).
\end{lstlisting}

Interestingly, we can use our algorithm to convert
a program of the extended fragment into its equivalent \langH program.
This can be done by initializing the transformation process with
the top predicate (here \lstinline{conj3(P,Q,R,X)}).
% Of course, the resulting program will still contain free predicate variables
% (since that is the case of the initialization atom)
% but it will construct a valid \langH program (since all unnecessary complex
% arguments will be eliminated in the specialized program).

%The implementation is approximately 700 lines of Prolog code and is realized for
%the XSB system. The source code of the transformation, as well as the experimental
%setup and the experimental results are publicly
%available.~\footnote{cf. \url{http://bitbucket.org/antru/firstify}}

%%%%%%%%%%%%%%%%%%%%%%%%%%%%%%%%%%%%%%%%%%%%%%%%%%%%%%%%%%%%%%%%%%%%%%%%%%%%%%%
%%%%%%%%%%%%%%%%%%%%%%%%%%%%%%%%%%%%%%%%%%%%%%%%%%%%%%%%%%%%%%%%%%%%%%%%%%%%%%%

%%
\section{Experiments}
\label{sec:experiments}

In this section we present some experiments to illustrate that our
technique can lead to the improvement of the execution runtime of higher-order logic programs.

%\subsection{Experimental Setup}

We have tested our method with a set of benchmarks that include
the computation of the transitive closure of a chain of elements,
a $k$-ary disjunction and $k$-ary conjunction of $k$ relations (for $k=5,10$),
the computation of the shortest path programs of a directed acyclic graph
and a set of programs that deal with preference representation~\cite{holppref}.
The higher-order program is expressed in Hilog and executed using the Hilog module
of XSB. XSB essentially transforms Hilog programs into first-order programs
using the techniques and optimizations described in \cite{SagonasW95},
and it also uses an optimized WAM instruction set to efficiently execute Hilog. The
measuremenets obtained include these optimizations.
We compare them with the execution of the Prolog programs produced by the
our predicate specialization technique.
Apart from XSB~\footnote{version 3.7, cf. \url{http://xsb.sourceforge.net/}},
we also consider for the execution of the specialized program in other Prolog engines.
The Prolog engines that we use are SWI-Prolog~\footnote{version 7.2.3, cf. \url{http://www.swi-prolog.org/}},
and YAP~\footnote{version 6.2.2, cf. \url{http://www.dcc.fc.up.pt/~vsc/Yap/}}.
%
%
% \begin{enumerate}
%
% \item The transitive closure of a chain of elements.
%
% \item The computation of the $k$-ary disjunction $\bigcup_{i=1}^k r(i)$, where $k=5,10$.
%       We used two programs for the same computation.
%       The first one uses a non recursive computation of the form
%       $c_k = (\dots((r(1) \cup r(2)) \cup r(3)) \dots \cup r(k))$
%       while the second one uses a recursive definition of the form
%       $c_1 = r(1)\ ;\ c_k = r(k) \cup c_{k-1}$.
%
% \item The computation of $k$-ary conjunction $\bigcap_{i=1}^k r(i)$, same as before.
%
% \item A set of programs that deal with preference representation~\cite{holppref}.
%       The selection of the best elements ($winnow$), the selection of the $k$-best
%       elements ($wt(k)$), and finally the selection of the elements at level $k$
%       according to some preference relation ($w(k)$).
%
% \item Two shortest path programs over a directed acyclic graph. Their
%       implementation are based on $winnow$.
%
% \end{enumerate}
%
Every program is executed several times, each time with a
predefined set of facts.
% For example, $closure$ is executed 5 times, for $n=1000,2000,4000,6000,8000$ facts.
All data has been artificially generated.

%All experiments were performed on a Linux Desktop PC (Ubuntu 14.04 LTS)
%with Intel(R) Core(TM) i7-4790 CPU, 8 GB RAM.

In addition to the standard execution for Hilog and Prolog code, we also perform a
\term{tabled execution} of both the higher-order and the first-order programs
in XSB. Tabling is a standard optimization technique that is widely used in
Prolog systems. In this optimization, a re-evaluation of a \term{tabled predicate}
is avoided by memoizing (ie. remembering) its answers. The XSB system is known
for its elaborate and efficient implementation of tabling for first-order logic
programs. For higher-order Hilog programs however, XSB's tabling mechanism may
not be as effective as it is for first-order ones. The reason is that in order to
table any Hilog predicate one has to table all Hilog code. This may lead to high
memory consumption, and can be problematic for large-scale program development.
We decided to table all predicates of the first-order programs
as well, despite the fact that it might have been possible to make a more
efficient use of tabling in this case. The idea behind this decision is to
enable us to draw a fair comparison between tabled Hilog and tabled Prolog.

%\subsection{Results}

\begin{table}[t]
  \caption{Experiment results. All execution times are in seconds.}
  %\scriptsize
  \centering
  \label{tab:results}
  \begin{tabular}{lrrrrrrrrl}
    \toprule
      & \multicolumn{1}{c}{hilog}
      & \multicolumn{3}{c}{prolog}
      & \multicolumn{1}{c}{hilog}
      & \multicolumn{1}{c}{prolog}
      & \multicolumn{3}{c}{program size}
    \\
        \multicolumn{1}{l}{program}
      & \multicolumn{1}{c}{xsb}
      & \multicolumn{1}{c}{xsb}
      & \multicolumn{1}{c}{swi}
      & \multicolumn{1}{c}{yap}
      & \multicolumn{1}{c}{xsb+tab.}
      & \multicolumn{1}{c}{xsb+tab.}
      & \multicolumn{1}{c}{h.o.}
      & \multicolumn{1}{c}{f.o.}
      & \multicolumn{1}{c}{facts}
    \\
    \cmidrule(r){1-1} \cmidrule(l){2-2} \cmidrule(l){3-5} \cmidrule(l){6-6} \cmidrule(l){7-7} \cmidrule(l){8-10}

    \textsf{closure}        & 1744.829 &  17.426 &  15.813 &  8.782 & 16.980 & 17.067 &  3 &  3 & 1000-8000 \\
    \textsf{closure\_1000}  &   12.132 &   0.801 &   0.609 &  0.372 &  0.872 &  0.672 &  3 &  3 & 1000 \\
    \textsf{closure\_2000}  &   91.284 &   2.884 &   2.644 &  1.332 &  2.944 &  3.004 &  3 &  3 & 2000 \\
    \textsf{closure\_4000}  &  709.356 &  11.336 &  10.918 &  5.464 & 10.812 & 11.076 &  3 &  3 & 4000 \\
    \textsf{closure\_6000}  & 2365.728 &  25.536 &  23.459 & 13.532 & 25.236 & 25.548 &  3 &  3 & 6000 \\
    \textsf{closure\_8000}  & 5545.644 &  46.576 &  41.433 & 23.208 & 45.036 & 45.036 &  3 &  3 & 8000 \\

    \cmidrule(r){1-1} \cmidrule(l){2-2} \cmidrule(l){3-5} \cmidrule(l){6-6} \cmidrule(l){7-7} \cmidrule(l){8-10}

    \textsf{conj5}          &    9.887 &   1.090 &   0.026 &  0.010 &  2.918 &  0.571 &  3 &  6 & 1000-8000 \\
    \textsf{genconj(5)}     &    9.921 &   1.101 &   0.028 &  0.011 &  2.031 &  0.573 &  4 &  4 & 1000-8000 \\
    \textsf{conj10}         &   21.676 &   2.414 &   0.023 &  0.015 & 11.741 &  1.276 &  3 & 11 & 1000-8000 \\
    \textsf{genconj(10)}    &   21.580 &   2.415 &   0.039 &  0.013 &  9.618 &  1.275 &  4 &  4 & 1000-8000 \\

    \cmidrule(r){1-1} \cmidrule(l){2-2} \cmidrule(l){3-5} \cmidrule(l){6-6} \cmidrule(l){7-7} \cmidrule(l){8-10}

    \textsf{union5}         &    0.035 &   0.028 &   0.030 &  0.023 &  0.037 &  0.038 &  4 & 10 & 1000-8000 \\
    \textsf{genunion(5)}    &    0.034 &   0.030 &   0.025 &  0.021 &  0.050 &  0.042 &  5 &  5 & 1000-8000 \\
    \textsf{union10}        &    0.063 &   0.062 &   0.046 &  0.036 &  0.075 &  0.065 &  4 & 20 & 1000-8000 \\
    \textsf{genunion(10)}   &    0.062 &   0.079 &   0.054 &  0.035 &  0.091 &  0.104 &  5 &  5 & 1000-8000 \\

    \cmidrule(r){1-1} \cmidrule(l){2-2} \cmidrule(l){3-5} \cmidrule(l){6-6} \cmidrule(l){7-7} \cmidrule(l){8-10}

    \textsf{path\_dag}      &  971.326 & 679.557 & 975.027 & 54.156 &  0.001 &  0.001 &  6 &  6 & 10-80 \\
    \textsf{path\_naive}    &    5.725 &   4.248 &   6.661 &  0.407 &  0.021 &  0.016 &  6 &  6 & 10-80 \\

    \cmidrule(r){1-1} \cmidrule(l){2-2} \cmidrule(l){3-5} \cmidrule(l){6-6} \cmidrule(l){7-7} \cmidrule(l){8-10}

    \textsf{winnow}         &    0.147 &   0.130 &   0.117 &  0.039 &  1.107 &  1.115 &  3 &  3 & 1000-10000 \\
    \textsf{w(2)}           &    3.920 &   3.257 &   3.844 &  0.527 &  0.168 &  0.213 & 10 & 12 & 100-2000 \\
    \textsf{w(3)}           &  129.457 & 107.183 & 122.556 & 21.103 &  0.119 &  0.123 & 10 & 12 & 100-2000 \\
    \textsf{wt(2)}          &    4.146 &   3.288 &   3.857 &  0.530 &  0.144 &  0.219 & 11 & 13 & 100-2000 \\
    \textsf{wt(3)}          &  130.540 & 108.048 & 126.876 & 21.360 &  0.100 &  0.119 & 11 & 13 & 100-2000 \\
  \bottomrule
  \end{tabular}
\end{table}

Table~\ref{tab:results} summarizes the experimental results. The
average execution time is depicted in seconds for each program and for each engine. The
execution time is measured using the standard {\tt time/1} predicate.
Apart from the execution time, the table also contains the number of the (non-fact)
clauses of the original higher-order program, the number of the (non-fact) clauses
of the resulting first-order program after the transformation, and the ranges of
the number of the corresponding facts. We do not show the runtime of each
transformation from the higher-order
to first-order since the execution of process was negligible
(e.g. less than 0.01 seconds in all cases).

Firstly, we observe that the first-order programs are in general much faster
than the higher-order ones. Even in the context of XSB which offers a native
support of Hilog, the Prolog code is in almost all cases faster than the Hilog
code. Especially in the transitive closure and the $k$-ary conjunction, we have
an improvement by one or more of orders of magnitude. In most programs in our
experiment, we noticed that the ratio between the execution time of Prolog code
and the execution time of Hilog code does not change much if we increase the
number of facts, with the exception of the transitive closure benchmark,
in which the more we increase the number of facts, the more this ratio decreases.
The most important advantage of executing standard Prolog though, is that it
allows us to choose from a wide range of available Prolog engines.
From the three Prolog engines that we used, YAP is the most performant one. Therefore,
we can get a further decrease in execution times by simply choosing a different
Prolog engine, a fact that is not possible if we want to execute Hilog code
directly.

As we stated earlier, tabling is another standard optimization technique that is
widely used in Prolog systems. Tabling was very effective in many cases in the
experiment, especially in the preference operations (\textsf{winnow}, \textsf{w} and \textsf{wt}) and
in the path programs (notice the dramatic decrease in the execution times for
the \textsf{path\_dag} benchmark). It seems that the performance of this
optimization offers the same performance gain for both Hilog and Prolog code,
since the execution times are in most cases similar. A notable exception is that
of the $k$-ary conjunction benchmark, in which the tabled Prolog code is 5 to 10
times faster than that of the tabled Hilog code. Also, the fact that we table
all Hilog and Prolog code did not have much negative effect in our experiment
after all, because (with the sole exception of the winnow benchmark) the tabled
executions are not slower than their non-tabled counterparts.

Finally, consider the programs that deal with the $k$-ary conjunction and
disjunction, i.e. the pairs \textsf{conj5} -- \textsf{genconj(5)}, \textsf{conj10 -- genconj(10)},
\textsf{union5 -- genunion(5)} and \textsf{union10 -- genunion(10)}.
Both programs of each of these pairs are making the same computation, with the
former expressed in a non-recursive way and the latter in a recursive way. These
programs differ also in the size of their first order counterparts.
The first-order form of the non-recursive version has more clauses
than the first-order form of the recursive version. We observe that
both the higher-order and the first-order versions of the same computation have
similar execution times, even though the first-order versions have different
numbers of clauses. As a result, an increase on the size of the
first-order program did not produce any overhead in the overall
program execution time.

%%%%%%%%%%%%%%%%%%%%%%%%%%%%%%%%%%%%%%%%%%%%%%%%%%%%%%%%%%%%%%%%%%%%%%%%%%%%%%%

%%
\section{Related Work}
\label{sec:related}
%%

% Partial Evaluation
The proposed predicate specialization is closely connected with related work on
\term{partial evaluation} of logic programs~\cite{LloydS91,Gallagher93,Leuschel98}.
More specifically, the proposed technique is a special form of partial evaluation
which targets higher-order arguments and uses a simple one-step unfolding rule
to propagate the constant higher-order arguments without changing the structure
of the original program. Consequently, first-order programs remain unchanged.
To the extend of our knowledge, partial evaluation techniques have not been
previously applied directly to higher-order logic programming with the purpose
to produce a simpler first-order program.

% Defunctionalization
%Techniques that aim to remove higher-order parameters have not been applied in
%the higher-order logic programming domain (e.g. in
%languages such as $\lambda$-Prolog~\cite{lambdaprolog} and Hopes~\cite{hopes}).
Other techniques, however, have been proposed that focus on the removal of higher-order
parameters in logic programs. Warren, in one of the early papers that tackle similar issues \cite{warren1981higher},
proposed that simple higher-order structures are non-essential and can be easily
encoded as first-order logic programs. The key idea is that every higher-order
argument in the program can be encoded as a symbol utilizing its name and a special
{\tt apply} predicate should be introduced to distinguish between different higher-order
calls.
A very similar approach has been employed in Hilog~\cite{hilog}; a language that
offers a \term{higher-order syntax} with first-order semantics. A Hilog program
is transformed into an equivalent first-order one using a transformation similar
to Warren's technique~\cite{warren1981higher}.
Actually, these techniques are
closely related to Reynolds' \term{defunctionalization}~\cite{Reynolds72} that
has been originally proposed to remove higher-order arguments in functional programs.
These techniques are designed to be applied in arbitrary programs in comparison
to our approach. In order to achieve this they require data structures in the
resulting program. However, on a theoretical view this imposes the requirement that the target
language should support data structures even if the source language does not support that.
This is apparent when considering Datalog; transforming a higher-order Datalog program
will result into a first-order Prolog program. On a more practical point, the
generic data structures introduced during the defunctionalization render the
efficient implementation of these programs challenging. The wrapping of the
higher-order calls with the generic apply predicate makes it cumbersome to
utilize the optimizations in first-order programs such as indexing and tabling.
In comparison, our technique produce more natural programs that do not suffer
for this phenomenon. Moreover, it does not introduce any data structures and
as a result a higher-order Datalog program will be transformed into a first-order one
amenable to more efficient implementation.

% Defunctionalization optimizations (XSB and Partial Evaluation)
In order to remedy the shortcomings of defunctionalization
there have been proposed some techniques to improve the performance of the transformed programs.
\citet{SagonasW95} proposed a compile-time optimization of the classical Hilog
encoding that eliminates some partial applications using a family of apply predicates
thus increasing the number of the predicates in the encoded program, which leads
to a more efficient execution. The original first-order encoding of Hilog as
well as this optimization are included in the XSB system~\cite{xsb}.
In the context of functional-logic programming,
there exist some mixed approaches that consider defunctionalization
together with partial evaluation for functional-logic programs~\cite{AlbertHV02,RamosSV05},
where a partial evaluation process is applied in a defunctionalized
functional-logic program.
Even though these approaches can usually offer a substantial performance improvement,
the resulting programs still use a Reynolds' style encoding; for instance,
the performance gain of the optimizations offered by XSB is not sufficient
when compared to the technique presented in this paper,
as presented in Section~\ref{sec:experiments}.

% Even though these approaches offer a substantial
% performance improvement in some cases, the resulting programs still use a
% Reynolds' style encoding which can be problematic in some cases.
% For example, in the case of XSB, our resulting first-order programs are generally
% more natural and in some cases can be executed more efficiently than the
% resulting first-order programs obtained by XSB.
% Secondly, in the case of XSB the transformations are executed
% at compile-time, while in our approach the transformation operates by using
% a given specialization goal in order to produce an efficient first-order program
% for this specific goal.

% Higher-order removal in Functional

The process of eliminating higher-order functions is being studied extensively in
the functional programming domain. Apart from defunctionalization, there exist some approaches that
do not introduce additional data structures while removing higher-order functions.
These techniques include the \term{higher-order removal} method of \citet{ChinD96},
the \term{firstification} technique of \citet{Nelan92} and the \term{firstify}
algorithm of \citet{MitchellR09}. The removal of higher-order values here is
achieved without introducing additional data structures, so the practical outcome
is that the resulting programs can be executed in a more efficient way than the
original ones. The basic operation of these transformation methods is
\term{function specialization}, which involves generating a new function in
which the function-type arguments of the original definition are eliminated.
A predicate specialization operation is also the core operation in our approach, so in
this point these approaches are similar to ours. The remaining operations that
can be found in those approaches (e.g. simplification rules, inlining,
eta-abstractions etc.), are either inapplicable to our domain or not needed for
our program transformation. Contrary to Reynolds' defunctionalization, these
higher-order removal techniques~\cite{ChinD96,MitchellR09,Nelan92} are not
complete, meaning that they do not remove all higher-order values from a
functional program, and therefore the resulting programs are not always first
order. This phenomenon would happen in our case as well if we considered the full
power of higher-order programming.
However, because of the fact that we focus on a smaller but
still useful class of higher-order logic programs, we are sure that the output
of our transformation technique will produce a valid first-order program for
every program that belongs to our fragment.

%%%%%%%%%%%%%%%%%%%%%%%%%%%%%%%%%%%%%%%%%%%%%%%%%%%%%%%%%%%%%%%%%%%%%%%%%%%%%%%
%%%%%%%%%%%%%%%%%%%%%%%%%%%%%%%%%%%%%%%%%%%%%%%%%%%%%%%%%%%%%%%%%%%%%%%%%%%%%%%
%%
\section{Conclusions and Future Work}
\label{sec:future}

In this paper we presented a program transformation technique that reduces
higher-order programs into first-order ones through argument specialization. The
transformation does not introduce additional data structures and therefore the
resulting programs can be executed efficiently in any standard Prolog system.
We do not consider the full power of higher-order logic programming, but we
focus on a modest but useful class of programs; in these programs
we do not allow partial applications or existential predicate variables in
the body of a clause.

In our actual implementation we considered a slightly broader class than the
fragment discussed before; we allowed a limited use of partial applications in
the case of predicates that do not belong to the same cycle in the predicate
dependency graph. This extension however does not increase the expressive power
of the language. An interesting open question that arises is whether this
technique can be used as a first-order reduction method only for programs that
belong to our fragment (or a fragment that have the same expressive power as ours)
or if it can be used for a wider class of programs that are more expressive
than our fragment. Moreover, any expansion of the supported class would be
desirable, even if it has the same expressivity as our current fragment.

Until now, we have used and evaluated our transformation technique only as an
optimization method for performance improvement. However, in the functional
programming domain, such techniques have been used in additional applications,
such as program analysis~\cite{MitchellR09} and implementation of debuggers~\cite{PopeN02}.
Therefore, an interesting aspect for future investigation would be the search of
similar or completely new applications of our higher-order removal technique in
the logic programming domain.

% As we stated and in the introduction, the practice in the area of the functional
% programming indicates that it is not possible to remove all higher-order
% elements for every program without introducing additional data structures. We
% have no reasons to believe that this will not be the case in the higher-order
% logic programming case. So, an interesting aspect for future investigation is
% to determine the exact class of higher-order logic programs that can be
% transformed into a finite first-order one without the use of additional data
% structures.

\subsubsection*{Acknowledgements}
We would like to thank the anonymous reviewers for providing constructive comments
on our original submission.

\bibliographystyle{splncsnat}
\bibliography{firstify}

\begin{thebibliography}{21}
\providecommand{\natexlab}[1]{#1}
\providecommand{\url}[1]{\texttt{#1}}
\providecommand{\urlprefix}{}

\bibitem[{Albert et~al.(2002)Albert, Hanus, and Vidal}]{AlbertHV02}
Albert, E., Hanus, M., Vidal, G.: A practical partial evaluation scheme for
  multi-paradigm declarative languages.
\newblock Journal of Functional and Logic Programming 2002 (2002)

\bibitem[{Charalambidis et~al.(2013)Charalambidis, Handjopoulos, Rondogiannis,
  and Wadge}]{hopes}
Charalambidis, A., Handjopoulos, K., Rondogiannis, P., Wadge, W.W.: Extensional
  higher-order logic programming.
\newblock {ACM} Transactions on Computational Logic 14(3), 21 (2013)

\bibitem[{Charalambidis et~al.(2018)Charalambidis, Rondogiannis, and
  Troumpoukis}]{holppref}
Charalambidis, A., Rondogiannis, P., Troumpoukis, A.: Higher-order logic
  programming: An expressive language for representing qualitative preferences.
\newblock Science of Computer Programming 155, 173 -- 197 (2018)

\bibitem[{Chen et~al.(1993)Chen, Kifer, and Warren}]{hilog}
Chen, W., Kifer, M., Warren, D.S.: Hilog: A foundation for higher-order logic
  programming.
\newblock Journal of Logic Programming 15(3), 187--230 (1993)

\bibitem[{Chin and Darlington(1996)}]{ChinD96}
Chin, W., Darlington, J.: A higher-order removal method.
\newblock Lisp and Symbolic Computation 9(4), 287--322 (1996)

\bibitem[{Gallagher(1993)}]{Gallagher93}
Gallagher, J.P.: Tutorial on specialisation of logic programs.
\newblock In: Proceedings of the {ACM} {SIGPLAN} Symposium on Partial
  Evaluation and Semantics-Based Program Manipulation, PEPM'93, Copenhagen,
  Denmark, June 14-16, 1993. pp. 88--98 (1993)

\bibitem[{Jones(2001)}]{jones}
Jones, N.D.: The expressive power of higher-order types or, life without
  {CONS}.
\newblock J. Funct. Program. 11(1), 5--94 (2001)

\bibitem[{Jones et~al.(1993)Jones, Gomard, and Sestoft}]{jonesetal93}
Jones, N.D., Gomard, C.K., Sestoft, P.: Partial evaluation and automatic
  program generation.
\newblock Prentice Hall (1993)

\bibitem[{Leuschel(1998)}]{Leuschel98}
Leuschel, M.: Logic program specialisation.
\newblock In: Partial Evaluation - Practice and Theory, {DIKU} 1998
  International Summer School, Copenhagen, Denmark, June 29 - July 10, 1998.
  pp. 155--188 (1998)

\bibitem[{Leuschel and Vidal(2014)}]{LeuschelVidal14}
Leuschel, M., Vidal, G.: Fast offline partial evaluation of logic programs.
\newblock Inf. Comput. 235, 70--97 (2014)

\bibitem[{Lloyd and Shepherdson(1991)}]{LloydS91}
Lloyd, J.W., Shepherdson, J.C.: Partial evaluation in logic programming.
\newblock J. Log. Program. 11(3{\&}4), 217--242 (1991)

\bibitem[{Miller and Nadathur(2012)}]{lambdaprolog}
Miller, D., Nadathur, G.: Programming with Higher-Order Logic.
\newblock Cambridge University Press, New York, NY, USA, 1st edn. (2012)

\bibitem[{Mitchell and Runciman(2009)}]{MitchellR09}
Mitchell, N., Runciman, C.: Losing functions without gaining data: another look
  at defunctionalisation.
\newblock In: Proceedings of the 2nd {ACM} {SIGPLAN} Symposium on Haskell,
  Haskell 2009, Edinburgh, Scotland, UK, 3 September 2009. pp. 13--24 (2009)

\bibitem[{Nelan(1991)}]{Nelan92}
Nelan, G.: Firstification.
\newblock Ph.D. thesis, Arizona State University (1991)

\bibitem[{Pope and Naish(2002)}]{PopeN02}
Pope, B.J., Naish, L.: Specialisation of higher-order functions for debugging.
\newblock Electr. Notes Theor. Comput. Sci. 64, 277--291 (2002)

\bibitem[{Ramos et~al.(2005)Ramos, Silva, and Vidal}]{RamosSV05}
Ramos, J.G., Silva, J., Vidal, G.: Fast narrowing-driven partial evaluation for
  inductively sequential programs.
\newblock In: Proceedings of the 10th {ACM} {SIGPLAN} International Conference
  on Functional Programming, {ICFP} 2005, Tallinn, Estonia, September 26-28,
  2005. pp. 228--239 (2005)

\bibitem[{Reynolds(1972)}]{Reynolds72}
Reynolds, J.C.: Definitional interpreters for higher-order programming
  languages.
\newblock In: Proc. of the 25th {ACM} Nat. Conf. pp. 717--740. ACM (1972)

\bibitem[{Sagonas and Warren(1995)}]{SagonasW95}
Sagonas, K., Warren, D.S.: Efficient execution of hilog in wam-based prolog
  implementations.
\newblock In: Proceedings of the 12th International Conference on Logic
  Programming, Tokyo, Japan, June 13-16, 1995. pp. 349--363 (1995)

\bibitem[{Shepherdson(1992)}]{Shepherdson92}
Shepherdson, J.C.: Unfold/fold transformations of logic programs.
\newblock Mathematical Structures in Computer Science 2(2), 143--157 (1992)

\bibitem[{Swift and Warren(2012)}]{xsb}
Swift, T., Warren, D.S.: {XSB:} extending prolog with tabled logic programming.
\newblock {TPLP} 12(1-2), 157--187 (2012)

\bibitem[{Warren(1982)}]{warren1981higher}
Warren, D.H.: Higher-order extensions to prolog-are they needed.
\newblock Machine Intelligence 10, 441--454 (1982)

\end{thebibliography}
\end{document}